\documentclass{sig-alternate-05-2015}

\usepackage{amssymb}
\usepackage{graphicx}
\graphicspath{{}}
\DeclareGraphicsExtensions{.jpg}
\usepackage{caption}
\usepackage{array}
\usepackage{stfloats}
\usepackage{multirow}
\usepackage[export]{adjustbox}

\usepackage{caption,subfig}

\usepackage{cite}
\usepackage{comment}

\usepackage{amsmath}

\makeatletter
\def\@copyrightspace{\relax}
\makeatother

\def\sharedaffiliation{%
\end{tabular}
\begin{tabular}{c}}
\begin{document}

\title{Identifying the potential of Near Data Computing for Apache Spark}

\numberofauthors{4} 
\author{
\alignauthor
        Ahsan Javed Awan\\
        \email{ajawan@kth.se}
% 2nd. author
\alignauthor
        Mats Brorsson\\
        \email{matsbror@kth.se}
% 3rd. author
\and
\alignauthor 
        Vladimir Vlassov\\
        \email{vladv@kth.se}
        % 4th author
\alignauthor 
       Eduard Ayguade\textbf{*}\\
       \email{eduard.ayguade@bsc.es}
\sharedaffiliation
       \affaddr{KTH Royal Institute of Technology}\\
        \affaddr{Department of Software and Computer Systems}\\
\sharedaffiliation
        \affaddr{\textbf{*}Barcelona Super Computing Center(BSC)}\\
        \affaddr{\textbf{*}Technical University of Catalunya(UPC)}\\        
}

\maketitle

\begin{abstract}
While cluster computing frameworks are continuously evolving to provide real-time data analysis capabilities, Apache Spark has managed to be at the forefront of big data analytics for being a unified framework for both,  batch and stream data processing. There is also a renewed interest is Near Data Computing (NDC) due to technological advancement in the last decade. However, it is not known if NDC architectures can improve the performance of big data processing frameworks such as Apache Spark. In this position paper, we hypothesize in favour of NDC architecture comprising programmable logic based hybrid 2D integrated processing-in-memory and in-storage processing for Apache Spark, by extensive profiling of Apache Spark based workloads on Ivy Bridge Server.

\end{abstract}

%\keywords{Processing in Memory, In-Storage Processing, Apache Spark}

\section{Introduction}
With a deluge in the volume and variety of data collecting, web enterprises (such as Yahoo, Facebook, and Google) run big data analytics applications using clusters of commodity servers. While cluster computing frameworks are continuously evolving to provide real-time data analysis capabilities, Apache Spark~\cite{Spark} has managed to be at the forefront of big data analytics for being a unified framework for SQL queries, machine learning algorithms, graph analysis and stream data processing. Recent studies on characterizing in-memory data analytics with Spark show that (i) in-memory data analytics are bound by the latency of frequent data accesses to DRAM~\cite{performance_spark} and (ii) their performance deteriorates severely as we enlarge the input data size due to significant wait time on I/O~\cite{performance_spark_volume}.

The concept of near-data computing (NDC) is regaining the attention of researchers partially because of technological advancement and partially because moving the compute closer to the data where it resides, can remove the performance bottlenecks of big data analysis workloads. The umbrella of NDC covers 2D-integrated Processing-In-Memory, 3D-stacked Processing-In-Memory (PIM) and In-Storage Processing (ISP). Existing studies show efficacy of  processing-in-memory (PIM) approach for simple map-reduce applications~\cite{pugsley2014ndc, islam2014improving}, graph analytics~\cite{ahn2015scalable, instruction_offloading}, machine learning applications~\cite{leebssync, bender2015k} and SQL queries~\cite{mirzadeh2015sort, xi2015beyond}. Researchers also show the potential of processing in non-volatile memories for I/O bound big data applications~\cite{ranganathan2011microprocessors, chang2012limits, wang2015propram}. However, it is not clear which aspect of NDC (high bandwidth, improved latency, reduction in data movement, etc..) will benefit state-of-art big data frameworks like Apache Spark. Before quantifying the performance gain achievable by NDC for Spark, it is pertinent to answer which form of NDC (PIM, ISP) would better suit Spark workloads?

To answer this, we characterize Apache Spark workloads into compute bound, memory bound and I/O bound. We use hardware performance counters to identify the memory bound applications and OS level metrics like CPU utilization, idle time and wait time on I/O to filter out the I/O bound applications in Apache Spark and position in favour of an NDC architecture with programmable logic based hybrid ISP and 2D integrated PIM.

\section{Background and Related Work}
\subsection{Spark}
Spark is a cluster computing framework that uses Resilient Distributed Datasets (RDDs), which are immutable collections of objects spread across a cluster. Spark programming model is based on higher-order functions that execute user-defined functions in parallel. These higher-order functions are of two types: ``Transformations'' and ``Actions''. Transformations are lazy operators that create new RDDs, whereas Actions launch a computation on RDDs and generate an output. When a user runs an action on an RDD, Spark first builds a DAG of stages from the RDD lineage graph. Next, it splits the DAG into stages that contain pipe-lined transformations with narrow dependencies. Further, it divides each stage into tasks, where a task is a combination of data and computation. Tasks are assigned to executor pool of threads. Spark executes all tasks within a stage before moving on to the next stage. Finally, once all jobs are completed, the results are saved to file systems. 

Spark MLlib is a machine learning library on top of Spark-core. GraphX enables graph-parallel computation in Spark. Spark SQL is a Spark module for structured data processing. It provides Spark with additional information about the structure of both the data and the computation being performed. This extra information is used to perform extra optimization. Spark Streaming provides a high-level abstraction called discretized stream or DStream, which represents a continuous stream of data. Internally, a DStream is represented as a sequence of RDDs. Spark streaming can receive input data streams from sources such as kafka. It then divides the data into batches, which are then processed by the Spark engine to generate the final stream of results in batches.

\subsection{Near Data Computing}
The umbrella of near-data computing covers both processing in memory and in-storage processing. A survey~\cite{siegl2016data} highlights historical achievements in technology that enables Processing-In-Memory (PIM) and various PIM architectures. It depicts PIM's advantages and challenges. Challenges of PIM architecture design are the cost-effective integration of logic and memory, unconventional programming models and lack of inter-operability with caches and virtual memory.

PIM approach can reduce the latency and energy consumption associated with moving data back-and-forth through the cache and memory hierarchy, as well as greatly increase memory bandwidth by sidestepping the conventional memory-package pin-count limitations. There exists a continuum of computing that can be embedded ``in memory''~\cite{loh2013processing}. This includes i) software transparent applications of logic in memory, ii) fixed function accelerators, iii) bounded operand PIM operations, which can be specified in a manner that is consistent with existing instruction-level memory operand formats, directly encoded in the opcode in the instruction set architecture, iv) compound PIM operations, which may access an arbitrary number of memory locations and perform number of different operations and v) fully programmable logic in memory, either a processor or re-configurable logic device.  

\subsection{Applications of PIM}
\textbf{PIM for Map-Reduce:} For Map-Reduce applications, prior studies~\cite{pugsley2014ndc,islam2014improving} propose simple processing cores in the logic layer of 3D-stacked memory devices to perform Map operations with efficient data access and without hitting the memory bandwidth wall. The reduce operations despite having random memory access patterns are performed on the central host processor.

\textbf{PIM for Graph Analytics:} The performance of graph analytics is bound by the inability of conventional processing systems to fully utilize the memory bandwidth and Ahn et al.~\cite{ahn2015scalable} propose in-order cores with graph processing specific prefetchers in the logic layer of 3D-stacked DRAM to fully utilize the memory bandwidth. Graph traversals are bounded by irregular memory access patterns of graph property and a study~\cite{instruction_offloading} proposes to offload the graph property to hybrid memory cube~\cite{hmc} (HMC) by utilizing the atomic requests described in HMC 2.0 specification (that is limited to only integer operations and one memory operand).

\textbf{PIM for Machine Learning:} Lee et al.~\cite{leebssync} use State Synchronous Parallel (SSP) model to evaluate asynchronous parallel machine learning workloads and observe that atomic operations are the hotspots and propose to offload them onto logic layers in 3D stacked memories. These atomic operations are overlapped with main computation to increase the execution efficiency. K-means, a popular machine learning algorithm, is shown to benefit from higher bandwidth achieved by physically bonding the memory to the package containing processing elements~\cite{bender2015k}. Another proposal~\cite{ncam} is to use content addressable memories with hamming distance units in the logic layer to minimize the impact of significant data movement in k-nearest neighbours.

\textbf{PIM for SQL queries:} Researchers also exploit PIM for SQL queries. The motivation for pushing select query down to memory is reduce data movement by pushing only relevant data up the memory hierarchy~\cite{xi2015beyond}. Join query can exploit 3D stacked PIM as it is characterized by irregular access patterns, but near-memory algorithms are required that consider data placement and communication cost and exploit locality with in one stack as much as possible~\cite{mirzadeh2015sort}

\textbf{PIM for Data Re-organization operations}: Another application of PIM is to accelerate data access and to help CPU cores to compute on complex linked data structures by efficiently packing them into the cache. Using strided DMA units, gather/scatter hardware and in-memory scratchpad buffers, the programmable near memory data rearrangement engines proposed in~\cite{gokhale2015near} perform fill and drain operations to gather the blocks of application data structures. 

\subsection{In-Storage Processing}
Ranganathan et al.~\cite{ranganathan2011microprocessors} propose nano-stores that co-locates processors and non-volatile memory on the same chip and connect to one another to form a large cluster for data-centric workloads that operate on more diverse data with I/O intensive, often random data access patterns and limited locality. Chang et al.~\cite{chang2012limits} examine the potential and limit of designs  that move compute in close proximity of NVM based data stores. The limit study demonstrates significant potential of this approach (3-162x improvement in energy-delay product) particularly for I/O intensive workloads. Wang et al.~\cite{wang2015propram} observe that NVM is often naturally incorporated with basic logic like data comparison write or flip-n-write module and exploit the existing resources inside memory chips to accelerate the key non-compute intensive functions of emerging big data applications.

\subsection{Big Data Frameworks and NDC}
Even though NDC seems promising for applications like map-reduce, machine learning algorithms, SQL queries and graph analytics,  but the existing literature lacks a study that identifies the potential of NDC for big data processing frameworks like Apache Spark, which run on top of Java Virtual Machine and use map-reduce programming model to enable machine learning, graph analysis  and SQL processing on batched and streaming data. One can argue that previous NDC proposals made only by studying the algorithms can be extrapolated to the big data frameworks but we refute the argument by stating that earlier proposal of using 3D-Stacked PIM for map reduce applications~\cite{pugsley2014ndc,islam2014improving} was motivated by the fact that the performance of map phase is limited by the memory bandwidth. Our experiments show that Apache Spark based map-reduce workloads don't fully utilize the available memory bandwidth. Prior work~\cite{awan2016node} also shows that high bandwidth memories are not needed for Apache Spark based workloads.

\section{Methodology}

Our study of identifying the potential of NDC to boost the performance of Spark workloads is based on matching the characteristics the Apache Spark based workloads to different forms of NDC (2D integrated PIM, 3D Stacked PIM, ISP)

\subsection{Workloads}

Our selection of benchmarks is inspired by~\cite{awan2016node}. Table~\ref{spark_workloads} shows the description of benchmarks. Big Data Generator Suite (BDGS), an open source tool is used to generate synthetic data sets based on raw data sets~\cite{BDGS}.

\begin{table}[!ht]
\renewcommand{\arraystretch}{1.3}
\centering
\caption{Spark Workloads}
\label{spark_workloads}
\resizebox{\columnwidth}{!}{
\begin{tabular}{l|l|l|l}
\hline
\multicolumn{1}{c|}{\textbf{\begin{tabular}[c]{@{}c@{}}Spark \\ Library\end{tabular}}} & \multicolumn{1}{c|}{\textbf{Workload}} & \multicolumn{1}{c|}{\textbf{Description}} & \multicolumn{1}{c}{\textbf{\begin{tabular}[c]{@{}c@{}}Input \\ data-sets\end{tabular}}} \\ \hline
\multirow{4}{*}{Spark Core} & \begin{tabular}[c]{@{}l@{}}Word Count \\ (Wc)\end{tabular} & counts the number of occurrence of each word in a text file & \multirow{2}{*}{\begin{tabular}[c]{@{}l@{}}Wikipedia \\ Entries\end{tabular}} \\ \cline{2-3}
 & Grep (Gp) & \begin{tabular}[c]{@{}l@{}}searches for the keyword The in a text file and filters out the\\ lines with matching strings to the output file\end{tabular} &  \\ \cline{2-4} 
 & Sort (So) & ranks records by their key & \begin{tabular}[c]{@{}l@{}}Numerical\\ Records\end{tabular} \\ \cline{2-4} 
 & \begin{tabular}[c]{@{}l@{}}NaiveBayes \\ (Nb)\end{tabular} & runs sentiment classification & \begin{tabular}[c]{@{}l@{}}Amazon \\ Movie \\ Reviews\end{tabular} \\ \hline
\multirow{6}{*}{Spark Mllib} & \begin{tabular}[c]{@{}l@{}}K-Means \\ (Km)\end{tabular} & \begin{tabular}[c]{@{}l@{}}uses  K-Means clustering algorithm from  Spark Mllib. \\ The benchmark is run for 4 iterations with 8 desired clusters\end{tabular} & \multirow{6}{*}{\begin{tabular}[c]{@{}l@{}}Numerical \\ Records\end{tabular}} \\ \cline{2-3}
% & \begin{tabular}[c]{@{}l@{}}Gaussian \\ (Gu)\end{tabular} & \begin{tabular}[c]{@{}l@{}}uses Gaussian clustering algorithm from Spark Mllib. \\ The benchmark is run for 10 iterations with 2 desired clusters\end{tabular} &  \\ \cline{2-3}
 & \begin{tabular}[c]{@{}l@{}}Sparse \\ NaiveBayes\\ (SNb)\end{tabular} & uses NaiveBayes classification alogrithm from Spark Mllib &  \\ \cline{2-3}
 & \begin{tabular}[c]{@{}l@{}}Support Vector\\ Machines (Svm)\end{tabular} & uses SVM classification alogrithm from Spark Mllib &  \\ \cline{2-3}
 & \begin{tabular}[c]{@{}l@{}}Logistic \\ Regression(Logr)\end{tabular} & uses Logistic Regression alogrithm from Spark Mllib &  \\ %\cline{2-3}
 %& Decision Tree (Dt) & uses Decision Tree alogrithm from Spark Mllib 
 \hline
\multirow{3}{*}{Graph X} & Page Rank (Pr) & \begin{tabular}[c]{@{}l@{}}measures the importance of each vertex in a graph.  \\ The benchmark is run for 20 iterations\end{tabular} & \multirow{3}{*}{\begin{tabular}[c]{@{}l@{}}Live \\ Journal\\ Graph\end{tabular}} \\ \cline{2-3}
 & \begin{tabular}[c]{@{}l@{}}Connected \\ Components (Cc)\end{tabular} & \begin{tabular}[c]{@{}l@{}}labels each connected component of the graph with the \\ ID of its lowest-numbered vertex\end{tabular} &  \\ \cline{2-3}
 & Triangles (Tr) & \begin{tabular}[c]{@{}l@{}}determines the number of triangles passing through \\ each vertex\end{tabular} &  \\ \hline
 \multirow{5}{*}{\begin{tabular}[c]{@{}l@{}}Spark\\ SQL\end{tabular}} & \begin{tabular}[c]{@{}l@{}}Aggregation \\ (SqlAg)\end{tabular} & \begin{tabular}[c]{@{}l@{}}implements aggregation query from BigdataBench \\ using DataFrame API\end{tabular} & \multirow{5}{*}{Tables} \\ \cline{2-3}
 & Join (SqlJo) & \begin{tabular}[c]{@{}l@{}}implements join query from BigdataBench \\ using DataFrame API\end{tabular} &  \\ \cline{2-3}
 & \begin{tabular}[c]{@{}l@{}}Difference \\ (Sql\_Diff)\end{tabular} & \begin{tabular}[c]{@{}l@{}}implements difference query from BigdataBench \\ using DataFrame API\end{tabular} &  \\ \cline{2-3}
 & \begin{tabular}[c]{@{}l@{}}Cross Product\\ (Sql\_Cro)\end{tabular} & \begin{tabular}[c]{@{}l@{}}implements cross product query from BigdataBench \\ using DataFrame API\end{tabular} &  \\ \cline{2-3}
 & \begin{tabular}[c]{@{}l@{}}Order By\\ (Sql\_Ord)\end{tabular} & \begin{tabular}[c]{@{}l@{}}implements order by  query from BigdataBench \\ using DataFrame API\end{tabular} &  \\ \hline
\multirow{3}{*}{\begin{tabular}[c]{@{}l@{}}Spark\\ Streaming\end{tabular}} & \begin{tabular}[c]{@{}l@{}}Windowed \\ Word Count \\ (WWc)\end{tabular} & \begin{tabular}[c]{@{}l@{}}generates every 10 seconds, word counts over the last \\ 30 sec of data received on a TCP socket every 2 sec.\end{tabular} & \multirow{3}{*}{\begin{tabular}[c]{@{}l@{}}Wikipedia \\ Entries\end{tabular}} \\ \cline{2-3}
 & \begin{tabular}[c]{@{}l@{}}Stateful Word\\  Count (StWc)\end{tabular} & \begin{tabular}[c]{@{}l@{}}counts words cumulatively in text received from the network\\  every sec starting with initial value of word count.\end{tabular} &  \\ \cline{2-3}
 & \begin{tabular}[c]{@{}l@{}}Network Word \\ Count (NWc)\end{tabular} & \begin{tabular}[c]{@{}l@{}}counts the number of words in the text, received from a data\\ server listening on a TCP socket every 2 sec and print the\\ counts on the screen. A data server is created by running \\ Netcat (a networking utility in Unix systems for creating \\ TCP/UDP connections)\end{tabular} &  \\ \hline
\end{tabular}
}
\end{table}

\subsection{System Configuration}

To  perform  our  measurements,  we  use  a  current  dual-socket  Intel  Ivy  Bridge  server  (IVB)  with  E5-2697  v2  processors, similar to what one would find in a datacenter. Table~\ref{hardware} shows details about our test machine. Hyper-Threading and Turbo-boost are disabled through BIOS during the experiments

\begin{table}[!ht]
\renewcommand{\arraystretch}{1.3}
\caption{Machine Details.}
\label{hardware}
\centering
\resizebox{\columnwidth}{!}{
\begin{tabular}{l|l|l}
\hline
\textbf{Component} & \multicolumn{2}{c}{\textbf{Details}} \\ \hline
Processor & \multicolumn{2}{l}{Intel Xeon E5-2697 V2, Ivy Bridge micro-architecture} \\ \hline
\multirow{6}{*}{} & Cores & 12 @ 2.7GHz (Turbo up 3.5GHz) \\ \cline{2-3} 
 & Threads & \begin{tabular}[c]{@{}l@{}}2 per Core (when Hyper-Threading \\ is enabled)\end{tabular} \\ \cline{2-3} 
 & Sockets & 2 \\ \cline{2-3} 
 & L1 Cache & \begin{tabular}[c]{@{}l@{}}32 KB for Instruction and \\ 32 KB for Data per Core\end{tabular} \\ \cline{2-3} 
 & L2 Cache & 256 KB per core \\ \cline{2-3} 
 & L3 Cache (LLC) & 30MB per Socket \\ \hline
Memory & \multicolumn{2}{l}{\begin{tabular}[c]{@{}l@{}}2 x 32GB, 4 DDR3 channels, Max BW 60GB/s\\ per Socket\end{tabular}} \\ \hline
OS & \multicolumn{2}{l}{Linux Kernel Version 2.6.32} \\ \hline
JVM & \multicolumn{2}{l}{Oracle Hotspot JDK 7u71} \\ \hline
Spark & \multicolumn{2}{l}{Version 1.5.0} \\ \hline
\end{tabular}
}
\end{table}

Table~\ref{parameters} lists the parameters of JVM and Spark after tuning. For our experiments, we configure Spark in local mode in which driver and executor run inside a single JVM. We use HotSpot JDK version 7u71 configured in server mode (64 bit) and use Parallel Scavenge (PS) and Parallel Mark Sweep for young and old generations respectively as recommended in~\cite{performance_spark_volume}. The heap size is chosen such that the memory consumed is within the system.
   
\begin{table}[]
\renewcommand{\arraystretch}{1.3}
\caption{Spark and JVM Parameters for Different Workloads.}
\label{parameters}
\centering
\resizebox{\columnwidth}{!}{
\begin{tabular}{l|clclc}
\hline
\multirow{2}{*}{\textbf{Parameters}} & \multicolumn{4}{c|}{\textbf{\begin{tabular}[c]{@{}c@{}}Batch \\ Processing\\ Workloads\end{tabular}}} & \multirow{2}{*}{\textbf{\begin{tabular}[c]{@{}c@{}}Stream \\ Processing \\ Workloads\end{tabular}}} \\ \cline{2-5}
 & \multicolumn{2}{c|}{\textbf{\begin{tabular}[c]{@{}c@{}}Spark-Core,\\ Spark-SQL\end{tabular}}} & \multicolumn{2}{c|}{\textbf{\begin{tabular}[c]{@{}c@{}}Spark Mllib, \\ Graph X\end{tabular}}} &  \\ \hline
spark.storage.memoryFraction & \multicolumn{2}{c|}{0.1} & \multicolumn{2}{c|}{0.6} & 0.4 \\ \hline
spark.shuffle.memoryFraction & \multicolumn{2}{c|}{0.7} & \multicolumn{2}{c|}{0.4} & 0.6 \\ \hline
spark.shuffle.consolidateFiles & \multicolumn{5}{c}{true} \\ \hline
spark.shuffle.compress & \multicolumn{5}{c}{true} \\ \hline
spark.shuffle.spill & \multicolumn{5}{c}{true} \\ \hline
spark.shuffle.spill.compress & \multicolumn{5}{c}{true} \\ \hline
spark.rdd.compress & \multicolumn{5}{c}{true} \\ \hline
spark.broadcast.compress & \multicolumn{5}{c}{true} \\ \hline
Heap Size (GB) & \multicolumn{5}{c}{50} \\ \hline
Old Generation Garbage Collector & \multicolumn{5}{c}{PS Mark Sweep} \\ \hline
Young Generation Garbage Collector & \multicolumn{5}{c}{PS Scavenge} \\ \hline
\end{tabular}
}
\end{table}

\subsection{Measurement Tools and Techniques}

We use Intel Vtune Amplifier~\cite{Vtune} to perform general micro-architecture exploration and to collect hardware performance counters. All measurement data are the average of three measure runs; Before each run, the buffer cache is cleared to avoid variation in the execution time of benchmarks. We use linux iotop command to measure the total disk bandwidth. To find sustained maximum bandwidth, we compile the OpenMP version of STREAM~\cite{stream_benchmark} using Intel's ICC compiler. We use linux top command in batch mode and monitor only java process of Spark to measure \%usr (percentage CPU used by user process) and \%io (percentage CPU waiting for I/O)

\section{Evaluation}

\subsection{The case of ISP for Spark}

\begin{figure*}[]
\centering
\subfloat[Average percentage CPU in user mode, wait on I/O and in idle state during the execution of Spark workloads]{\includegraphics[scale=0.40]{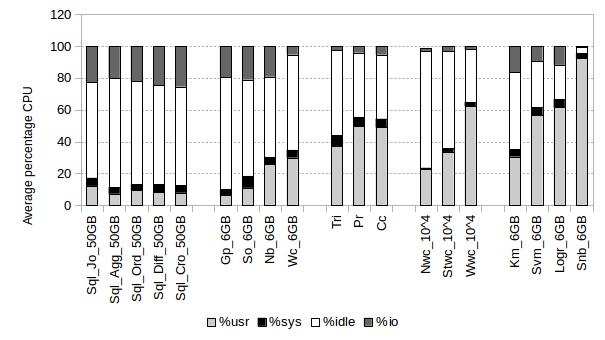}
\label{user_iowait}}
\subfloat[Spark workloads do not saturate the disk bandwidth]{\includegraphics[scale=0.40]{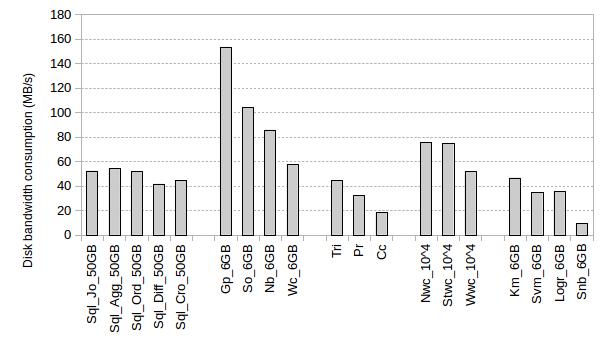}
\label{disk_bandwidth}}
\hfill
\subfloat[Spark workloads are DRAM bound]{\includegraphics[scale=0.40]{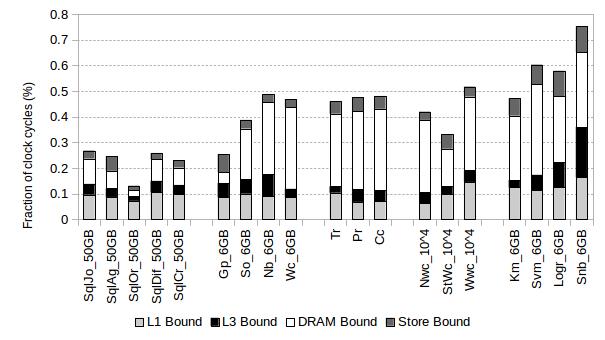}
\label{spark_membound}}
\subfloat[Spark workloads do not experience loaded latencies]{\includegraphics[scale=0.40]{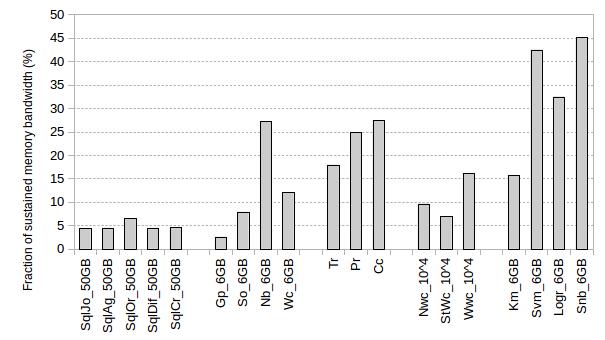}
\label{fraction_sustained}}
\hfill
\subfloat[Sql\_Join]{\includegraphics[scale=0.40]{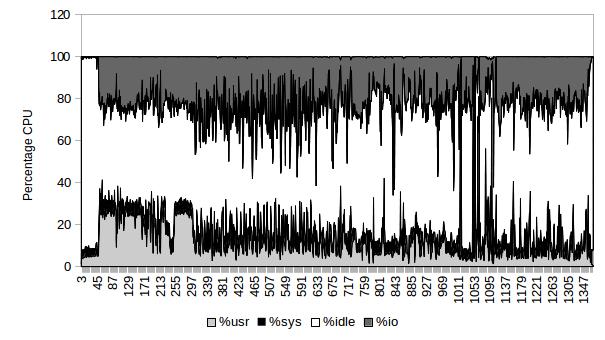}
\label{sqljo_usr_io}}
\subfloat[Windowed Word Count]{\includegraphics[scale=0.40]{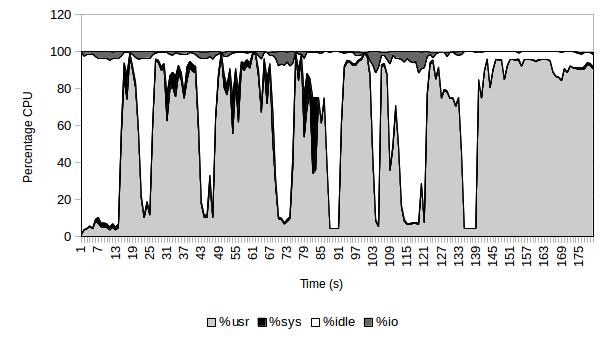}
\label{wwc_usr_io}}
\hfill
\subfloat[Page Rank]{\includegraphics[scale=0.40]{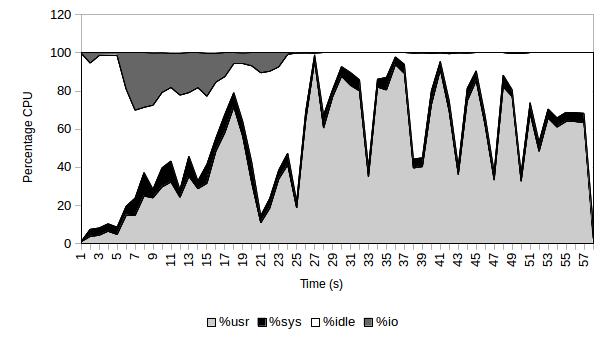}
\label{pr_usr_io}}
\subfloat[Kmeans]{\includegraphics[scale=0.40]{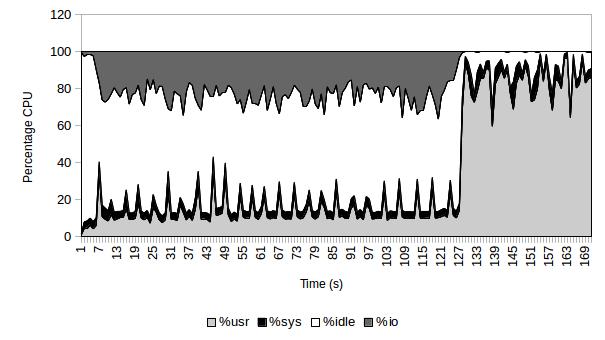}
\label{km_usr_io}}
\caption{Characterization of Spark workloads from NDC perspective}
\label{datavelocity}
\end{figure*}

Figure~\ref{disk_bandwidth} shows the average amount of data read from and written to the disk per second for different Spark workloads. The data reveal that on average across the workloads, total disk bandwidth consumption is 56 MB/s. The SATA HDD installed in the machine under test can support up to 164.5 MB/s of 128 KB sequential reads and writes. However, the average response time for 4 KB reads and writes are 1803.41ms and 1305.66ms respectively~\cite{Toshiba}. This implies that Spark workloads do not saturate the bandwidth of SATA HDD but the latency of I/O operations are detrimental to the performance of Spark workloads.

Figure~\ref{user_iowait} shows average percentage CPU, a) used by Spark java process, b) in system mode c) waiting for I/O and d) in idle state during the execution of different Spark workloads. Even though the number of Spark worker threads are equal to the number of CPUs available in the system, during the execution of Spark SQL queries, only 8.97\% CPUs are in user mode, 22.93\% CPUs are waiting for I/O and 63.52\% CPUs are in idle state.  We see similar characteristics for Grep and Sort. 

Grep, WordCount, Sort, NaiveBayes, Join, Aggregation, Cross Product, Difference and Orderby queries are all non iterative workloads, the data is read from and written to disk through out the execution period of workloads (see Figure~\ref{sqljo_usr_io}) and compute intensity varies from low to medium and the amount of data written to the disk also varies. For all these disk based workloads, we recommend in-storage processing. Since these workloads differ in the compute intensity, putting simple in-order cores would be less effective as compared to programmable logic, which can be programmed with workload specific hardware accelerators. Moreover, using hardware accelerators inside the NAND flash can free up the resources at the host CPU, which in turn can be used for other compute-intensive tasks.

\subsection{The case of PIM for Apache Spark}

When Graph-X workloads are run, 45.15\% CPUs are in the user mode, 3.98\% CPUs wait for I/O and 44.63\% CPUs are in the idle state. Pagerank, Connected Components and Triangle counting are iterative applications on graph data, which can easily fit in the main memory. All these workloads have a phase of heavy I/O with moderate cpu utilization followed by the phase of high cpu utilization and negligible I/O (see Figure~\ref{pr_usr_io}) These workloads is dominant by the second phase. 

During the execution of  stream processing workloads, 39.52\% CPUs are in the user mode, 2.29\% CPUs wait for I/O and 55.78\% CPUs are in the idle state. The wait time on I/O for stream processing workloads is negligible (see Figure~\ref{wwc_usr_io}) due the streaming nature of the workloads but the cpu utilization also varies from low to high.

For Spark MLlib workloads, the percentage of CPUs in user mode, waiting for I/O and in idle state are 60.27\%, 9.56\% and 25.48. SVM and Logistic Regression are phasic in terms of I/O. The training phase has significant I/O and also high CPU utilization, whereas the testing phase has negligible I/O and high CPU utilization because before the training starts, the input data is split into training and testing data and are cached in the memory. 

Since DRAM bound stalls are higher than L3 bound stalls and L1 bound stalls for most of the Graph-X, Spark Spark Streaming and Spark MLlib workloads (see Figure~\ref{spark_membound}), it means that CPUs are stalled waiting  for the data to be fetched from the main memory and not by the caches(for detailed analysis see~\cite{performance_spark_volume, performance_spark,awan2016micro}). So, instead of moving the data back and forth through the cache hierarchy in between the iterations, it would be beneficial to use programmable logic based processing-in-memory. As a result,  application specific hardware accelerators are brought closer to the data, which will reduce the data movement and improve the performance of Spark workloads.

\subsection{The case of 2D integrated PIM instead of 3D stacked PIM for Apache Spark}

According to Jacob et al.~\cite{jacob2009memory}, the bandwidth vs latency response curve for a system has three regions. For the first 40\% of the sustained bandwidth, the latency response is nearly constant. The average memory latency equals idle latency in the system and the system performance is unbounded by the memory bandwidth in the constant region. In between 40\% to 80\% of the sustained bandwidth, the average memory latency increases almost linearly due to contention overhead by numerous memory requests. The performance degradation of the system starts in this linear region. Between 80\% to 100\% of the sustained bandwidth, the memory latency can increase exponentially over the idle latency of DRAM system and the applications performance is limited by available memory bandwidth in this exponential region.

3D-Stacked PIM based on Hybrid Memory Cube (HMC) enables significantly more bandwidth between the memory banks and the compute units as compared to 2D integrated PIM, e.g. maximum theoretical bandwidth of 4 DDR3-1066 is 68.2 GB/s where as 4 HMC links provide 480 GB/s~\cite{radulovic2015another}. If the workload is operating in the exponential region on bandwidth vs latency curve of DDR3 based system, using HMC will move the workload to operate again in the constant region and average memory latency equals idle latency of the system. On the other hand, if the workloads are not bounded by the memory bandwidth, NDC architecture based on 3D-stacked PIM would not be able to fully utilize the excessive bandwidth and goal of reducing the data movement can be achieved instead by 2D integrated PIM.

Figure~\ref{fraction_sustained} shows the average bandwidth consumption as a fraction of sustained maximum bandwidth. The data reveal Spark workloads consume less than 40\% of sustained maximum bandwidth at 1866 data transfer rate and thus operate in the constant region. Awan et al.~\cite{awan2016node} study the bandwidth consumption of Spark workloads during the whole execution time of the workloads and show that even when the peak bandwidth utilization goes into the exponential region, it lasts only for a short period of time and thus, have a negligible impact on the performance. Thus we envision 2D integrated PIM instead of 3D stacked PIM for Apache Spark.

\subsection{The case of Hybrid 2D integrated PIM and ISP for Spark}

K-means is also  an iterative algorithm. It has two distinct phases (see Figure~\ref{km_usr_io}), heavy I/O phase followed by negligible I/O phase. The heavy IO phase has low  cpu utilization. This phase implements kmeans|| initialization method to assign initial values to the clusters. This phase can be mapped to hardware accelerators in the programmable logic inside the storage, where as the main clustering algorithm can be mapped to 2D integrated PIM.

\section{Conclusion}

We study the characteristics of Apache Spark workloads from the NDC perspective and and position ourselves as follows; i) Spark workloads, which are not iterative and have high ratio of \% cpu waiting for I/O to \% cpu in user mode like SQL queries, filter, word count and sort are ideal candidates for ISP, ii) Spark workloads, which have low ratio of \% cpu waiting for I/O to \% cpu in user mode like stream processing and iterative graph processing workloads are bound by latency of frequent accesses to DRAM and are ideal candidates for 2D integrated PIM,  iii) Spark workloads, which are iterative and have moderate ratio of \% cpu waiting for I/O to \%cpu in user mode like K-means, have both I/O bound and memory bound phases and hence will benefit from the combination of 2D integrated PIM and ISP and iv) to satisfy the varying compute demands of Spark workloads, we envision an NDC architecture with programmable logic based hybrid ISP and 2D integrated PIM.

Future work involves quantifying the performance gain for Spark workloads  achievable through programmable logic based ISP and 2D integrated PIM.

\section*{Acknowledgments}
This work is supported by  Erasmus Mundus Joint Doctorate in Distributed Computing (EMJD-DC) program  funded by the Education, Audiovisual and Culture Executive Agency (EACEA) of the European Commission. It is also supported by the Spanish Government through Programa Severo Ochoa (SEV-2015-0493), by the Spanish Ministry of Science and Technology through TIN2015-65316-P project and by the Generalitat de Catalunya (contract 2014-SGR-1051). We thank Moriyoshi Ohara for his comments on the first draft of the paper.

\bibliographystyle{acm}

\small{
\raggedright
\bibliography{thesis,references}

}

\balancecolumns
\end{document}